\documentclass[conference]{IEEEtran}
\IEEEoverridecommandlockouts
\usepackage{cite}
\usepackage{multirow}
\usepackage{amsmath,amssymb,amsfonts}
\usepackage{algorithmic}
\usepackage{graphicx}
\usepackage{textcomp}
\usepackage{xcolor}
\usepackage{subfigure}
\usepackage{caption}
\usepackage{subcaption}
\def\BibTeX{{\rm B\kern-.05em{\sc i\kern-.025em b}\kern-.08em
    T\kern-.1667em\lower.7ex\hbox{E}\kern-.125emX}}
\begin{document}

\title{
% Towards Identifying Cyberattack Intrusion In Controlled Dynamical Networks
Isolating Signatures of Cyberattacks under Stressed Grid Conditions
\thanks{This work was supported by the U.S. Department of Energy’s (DOE) Office of Cybersecurity, Energy Security, and Emergency Response (CESER) and performed at the Pacific Northwest National Laboratory (PNNL), operated for the U.S. DOE by Battelle Memorial Institute under Contract No. DE-AC05-76RL01830. Sanchita Ghosh is from Texas Tech University and this work is performed while she is an intern at PNNL.}
}

\author{\IEEEauthorblockN{Sanchita Ghosh, Syed Ahsan Raza Naqvi, Sai Pushpak Nandanoori, and Soumya Kundu}
\IEEEauthorblockA{\textit{Optimization \& Control Group} \\
\textit{Pacific Northwest National Laboratory, USA} \\
Emails: \{sanchita.ghosh\,,\,ahsan.raza\,,\,saipushpak.n\,,\,soumya.kundu\}@pnnl.gov}
% \and
% \IEEEauthorblockN{}
% \IEEEauthorblockA{\textit{Optimization \& Control Group)} \\
% \textit{Pacific Northwest National Laboratory}\\
% Richland, USA \\
% @pnnl.gov}
% \and
% \IEEEauthorblockN{}
% \IEEEauthorblockA{\textit{Optimization \& Control Group)} \\
% \textit{Pacific Northwest National Laboratory}\\
% Richland, USA \\
% @gmail.com}
% \and
% \IEEEauthorblockN{}
% \IEEEauthorblockA{\textit{Optimization \& Control Group)} \\
% \textit{Pacific Northwest National Laboratory}\\
% Richland, USA \\
% @pnnl.gov}
% \and
% \and
}

\maketitle

\begin{abstract}
In a controlled cyber-physical network, such as a power grid, any malicious data injection in the sensor measurements can lead to widespread impact due to the actions of the closed-loop controllers. While fast identification of the attack signatures is imperative for reliable operations, it is challenging to do so in a large dynamical network with tightly coupled nodes. A particularly challenging scenario arises when the cyberattacks are strategically launched during a grid stress condition, caused by non-malicious physical disturbances. 
% Reliable measurement data from sensors like Phasor Measurement Units (PMUs) are imperative in taking appropriate control actions to avoid any disruption of operation, particularly in the event of abrupt changes in load and generation. Unfortunately, attackers may corrupt the measurement data aiming to interrupt nominal operation leading to financial losses and equipment damages. Thus, it is crucial to detect these attacks and identify the corrupted sensor measurements to maintain the nominal operation. 
In this work, we propose an algorithmic framework -- based on Koopman mode (KM) decomposition -- for online identification and visualization of the cyberattack signatures in streaming time-series measurements from a power network. The KMs are capable of capturing the spatial embedding of both natural and anomalous modes of oscillations in the sensor measurements and thus revealing the specific influences of cyberattacks, even under existing non-malicious grid stress events. Most importantly, it enables us to quantitatively compare the outcomes of different potential cyberattacks injected by an attacker. The performance of the proposed algorithmic framework is illustrated on the IEEE 68-bus test system using synthetic attack scenarios. Such knowledge regarding the detection of various cyberattacks will enable us to devise appropriate diagnostic scheme while considering varied constraints arising from different attacks. 
\end{abstract}


\section{Introduction}
% Real-time operational control of  complex dynamical cyber-physical networks relies heavily on accurate sensor measurements received via communication networks \cite{wen2017complex}. Modernized power grid is an example of similar dynamical networks where cloud-based controllers such as automatic generation control (AGC) utilizes measurement data from advance sensors including phasor measurement units (PMUs) to maintain grid stability vua real-time control actions.

Due to the proliferation of advanced sensing technologies, including phasor measurement units (PMUs), the modern power grid can monitor and promptly control the power supply and demand. However, this increased dependence on monitoring and control for reliable operation also renders the grid susceptible to adversarial actions. A recent US Department of Energy report \cite{us2022cybersecurity} identified increasing cloud-based communication and control systems, alongside fast-evolving ransomware threats, and the convergence of information technology (IT) and operations technology (OT) systems, as some of the emerging cybersecurity challenges faced by the power grid today. According to a recent webcast by NERC, which was reported by Reuters \cite{Reuters_article}, the number of susceptible points in the power system is increasing by about 60 per day, and due to the increased geopolitical conflicts, there is a dramatic increase in the number of cyber attacks on the grid. Therefore, it is imperative to detect the cyber-attacks on the power grid at the earliest to avoid disruptions in the operations.

% Despite the development of bad data detectors to identify such intrusions, cyber adversaries have continued to adapt and their attacks have become increasingly sophisticated to inflict maximal damage on the power system. A few reported real-world scenarios of successful cyber attacks that caused disruptions to the power system include the Stuxnet \cite{karnouskos2011stuxnet}, the dragonfly \cite{symantec2017dragonfly}, and the 2015 cyber-attacks on the Ukrainian Grid \cite{case2016analysis}. 
% % According to a recent webcast by NERC, which was reported by Reuters \cite{Reuters_article}, the number of susceptible points in the power system is increasing by about 60 per day, and due to the increased geopolitical conflicts, there is a dramatic increase in the number of cyber attacks on the grid. Therefore, it is imperative to detect the cyber-attacks on the power grid at the earliest to avoid disruptions in the operations. 
% In this work, we focus on a class of cyber-attacks, namely the false data injection (FDI) attacks on PMUs that result in unwarranted control inputs from the automated controls that rely on those PMU measurements. PMUs are GPS time-stamped measurements that typically report 20-60 samples per second thus allowing visibility into the system's transient dynamics. 

In this work, we focus on a class of cyber-attacks, namely the false data injection (FDI) attacks on PMUs, that result in unwarranted control inputs from the automated controls that rely on those PMU measurements. Some existing grid cybersecurity solutions adopt a hybrid (combined IT/OT) intrusion detection approach using static, pre-determined, OT rules \cite{johnson2022,eos}. Rule-based cyber intrusion detection engines suffer from a lack of adaptability, especially under unforeseen grid stress events which tend to move the system away from normal (expected) operating conditions. Ongoing research into OT-based automated (non-rule-based) cyberattack detection has looked into different methods that: i) rely on physics-based models with an outlier detection algorithm (using state estimators) \cite{ghosal2018diagnosis,murguia2016characterization,huang2018online}; or ii)  adopt various machine learning and deep learning approaches \cite{abedi2023towards,kurt2018online,SPADES}, a comparison of some of which are presented in a recent article \cite{yin2024advancing}.

Early (and accurate) detection of FDI attacks on sensors is critical in ensuring reliable grid operations, since malicious sensors measurements may otherwise mislead the grid control actions and create system-wide adverse impact \cite{us2022cybersecurity}. Some of the major challenges towards accurate and early detection of cyberattacks stem from: nonlinearity of grid transient dynamics, and the presence of naturally caused (non-malicious) physical disturbances which may lead to stressed grid conditions. The work in \cite{ghosal2018diagnosis} relied on a linear physics-based model to develop an algorithm that distinguishes between the signatures of FDI attacks and non-malicious grid disturbances. Other research efforts have explored learning-based methods, e.g., supervised learning on large labeled datasets \cite{intriago2021online}, and non-supervised learning using auto-regressive models \cite{abedi2023towards}, for distinguishing between attacks and physical disturbance events. All these methods, however, assume linearity of the underlying system dynamics and do not extend to nonlinear and stressed grid conditions. In this work, we resolve these challenges by adopting a nonlinear modal analysis-based approach, via Koopman operators, which shows promise in online detection of cyberattacks in controlled dynamical systems \cite{nandanoori2020model,ghosh2024koopman}.

% Some of the drawbacks of the existing methods include challenges of modeling physics-based models, unforeseen changes and inaccurate estimates during transients, not having sufficient training data, and the need for computational resources. The attack detection can furthermore become challenging due to the non-linearity and complexity of power system dynamics which makes it difficult to set up a baseline for any conventional outlier-based detection algorithms. Additionally, due to the presence of closed-loop control actions, the FDI attack-based measurements propagate through the network, essentially implying that the FDI attack needs to be detected at the earliest.  
% The work in \cite{nandanoori2020model} developed an attack detection algorithm that addresses the above-mentioned challenges by relying only on sensor measurements from across a power network and using Koopman mode (KM) decomposition to reveal the presence of an attack in real-time. 
In this work, we develop an algorithmic framework, based on Koopman mode (KM)
decomposition, for online identification and visualization of the
cyberattack signatures in streaming time-series measurements, even under existing non-malicious grid stress event.
% extend the attack detection algorithm introduced in \cite{nandanoori2020model} to automatically detect the attack locations and also differentiate them from organic grid events, such as load changes. 
We achieve this in two steps: 

\begin{itemize}
    \item[(a)] Rigorous spatial and quantitative comparison between the KMs to reveal the outliers corresponding to an attack and natural events. The spatial distance between the KMs is computed using Kullback-Leibler (KL) distance to reveal the possible attack locations to the onlooking operators. 
    % (With the help of human-in-the-loop, the attack locations are now identified) 
    \item[(b)] A clustering algorithm to identify the attack locations from the spatial distance between the KMs without the need for the interpretation by the human operators. 
    % human-in-the-loop. 
\end{itemize}
\par In this paper, which is a work in progress, we present the first step (a) and are currently working on the second step (b). 

% \textbf{Challenges}
% \begin{itemize}
%     \item Localizing the source of attack in dynamical network strong connection among nodes such that the nodes mimic each other's dynamic signatures making it difficult to isolate the source at attacks.
%     \item Nonlinearity and complexity of dynamics behavior difficult to setup the baseline for any conventional outlier-based anomaly detection algorithms.
%     \item Presence of the closed loop control actions attack impact quickly propagates throughout the network.
% \end{itemize}

\begin{figure*}[ht]
    \centering
    \includegraphics[width=1\linewidth]{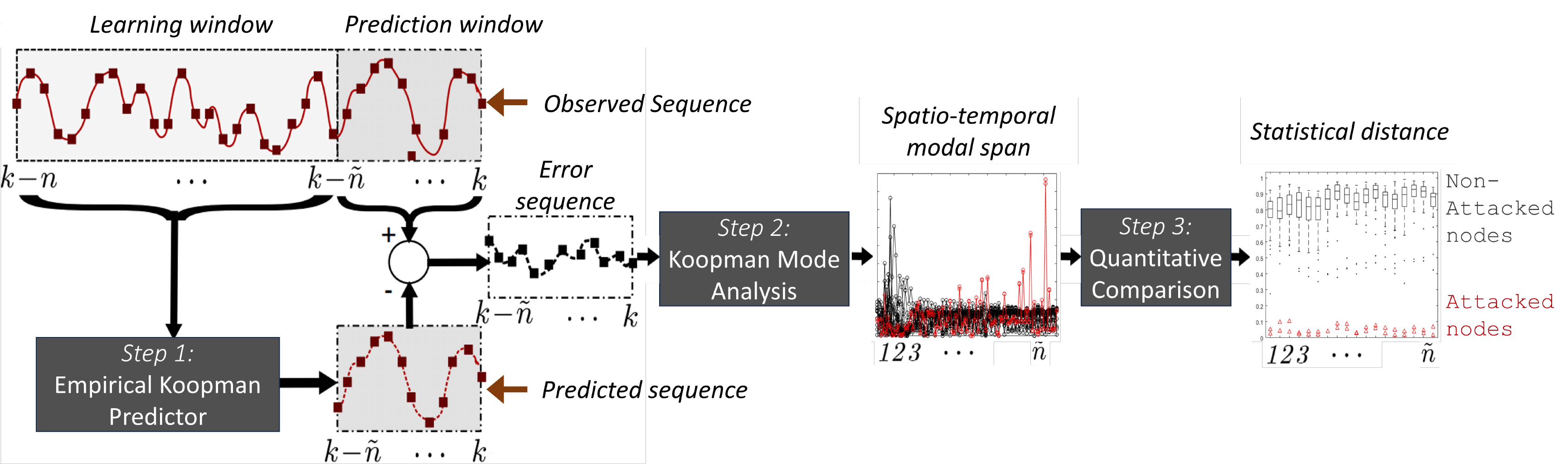}
    \caption{Step-wise schematic of the proposed quantitative comparison algorithm}
    \label{fig:methodology}
\end{figure*}

\section{Methodology}
In this work, we adopt the Koopman operator-based approach and follow the steps described in our previous paper \cite{nandanoori2020model}. We further investigate the specific influence of attacks on KMs to improve the reliability and efficiency of the detection algorithm against a varied range of attacks. In particular, we intend to identify the alterations in KMs due to an attack as opposed to a grid event. 
% while considering the transient dynamics of the power system during grid-stress operation. 
Fig \ref{fig:methodology} captures the steps of the proposed algorithm. We follow steps 1-2 to obtain individual KMs corresponding to each of the sensors for the error sequence over a prediction window (find details in \cite{nandanoori2020model}).

    \textbf{Step 1 (Empirical Koopman Predictor):} We utilize the time-series data from the PMU sensors to ascertain the empirical Koopman predictor over a learning window to generate the measurement prediction sequence for the prediction window. We obtain the error sequence from the predicted and the observed sequence for the corresponding prediction window.
    
    \textbf{Step 2 (Koopman Mode Decomposition):} We compute the KMs for the error sequence, which capture the latent spatio-temporal attack signatures in the presence of an attack.
    
   \textbf{Step 3 (Quantitative Comparison):} We utilize the KMs of the error sequence to conduct the 3 sub-steps described below.
   \begin{itemize}
       \item[(a)] We conduct a two-step normalization of the KMs. First, we normalize the KMs across the time points (i.\,e., each time point receives similar weights) and then, we further normalize the KMs across spatial nodes (i.\,e., each sensor node receives similar weights), such that we obtain:
       \begin{align*}
           \overline{v}_i&:=\left<\text{a row vector of normalized KMs of sensor-$i$}\right>
       \end{align*}       
       Note that such two-step normalization of the KMs corresponding to each sensor node over the prediction window allows the application of probabilistic distance measures as described below\footnote{After the normalization, the KMs corresponding to each sensor node over the prediction window represent a valid probability mass function, i.e., their values are non-negative and sum up to unity.}.
       \item[(b)] Next, to capture and quantify the changes in the distribution of the normalized KMs due to an attack, we utilize probabilistic distance measures such as KL divergence, and Jensen–Shannon (JS) divergence. Thus, we compute the distance from each KM to a centroid node with the distance measures. Here, we consider the spatial average KM (across all $p$ sensors) at each time point as the centroid node to measure the distances, i.e.,
       \begin{align*}
           \overline{v}_*&:={\sum}_{i=1}^p\overline{v}_i /p\qquad\left<\text{the centroid KM}\right>\\
           d(i,*)&:=\left<\text{distance between KM-$i$ and centroid KM}\right>
       \end{align*}
       \item[(c)]  Then, we define an exponential function to map the measured distance to KM $\Delta$-score (taking values between 0 and 1), with a lower $\Delta$-score implying higher distance from the centroid, and hence a likely attacked node,
       \begin{align*}
           \textbf{(KM $\Delta$-score)}\,~\,\Delta_i&\!:=\exp{\left(-\tau d(i,*)\right)}\,,\,~\text{for some $\tau\!>\!0$}
       \end{align*}
   \end{itemize}  Finally, we repeat steps 1-3 for a moving window to monitor the changes in KM $\Delta$-score distribution during grid-events and attacks with time. We note that the proposed algorithm quantifies the attack signature embedded in the KMs without prior attack knowledge. However, we utilize the available attack knowledge during the visualization of the embedded attack signature as described in following section.

\section{Results and Discussion}
% \begin{figure*}[t]
%         \centering
%        \hspace{-0.5in} \includegraphics[width=0.37\linewidth]{aspectRatio.png}\hspace{-0.27in}
%         \includegraphics[width=0.37\linewidth]{aspectRatio.png}\hspace{-0.26in}
%         \includegraphics[width=0.37\linewidth]{aspectRatio.png}\hspace{-0.54in}
% \end{figure*}
\subsection{System Description}
In this work, we utilize a MATLAB-based framework to simulate the dynamics for an IEEE 68-bus power system consisting of 16 synchronous generators (SGs) and equipped with sensors such as PMUs at each bus, measuring voltage angle, voltage magnitude, frequency and change in frequency.
% The SGs are modeled with 1-axis governor controls which acts as primary controls for frequency regulation. The system is assumed to be equipped with sensors such as PMUs at each bus, measuring voltage angle, voltage magnitude, frequency and change in frequency. 
The secondary control, automatic generation control (AGC) is also implemented using wide-area measurements such that the framework has the capability to model grid events, e.\,g., load changes, generator loss, line tripping, and faults. Moreover, the simulation runs with a simulation time step of $T = 0.05$s, i.\,e., we collect 20 time series data for each sensor per second which complies with real-world operations.
\subsection{Attack Scenario Description}
We consider an adversary injecting attack signals into the sensor measurements for voltage angles while the system undergoes some existing grid event. Thus we are focusing on the attacks that may stay hidden behind the transient dynamics of the system during a grid-stress operation. To realize this proposition, we consider that the system is in a steady state initially with a small load change occurring at load 1 from  $1$-$5\,$ s in every attack scenario.  As the system settles after this load change, we introduce a larger load change at bus 52 starting from the $15^{\text{th}}$ second. Then,  the attack starts at  $ t_1 = 20\,$s and continues until $t_2 = 40\,$s.
 The attack strategies studied in this work, as in \cite{yin2024advancing}, can be categorized into (i) additive attacks (‘poisoning’, ‘denial-of-service (DoS)’), (ii) scaling attacks (‘step’, ‘ramp’), and (iii) a combination of both (‘riding the wave (RTW)’). In general, we define the attacked measurement $\Tilde{\Phi}_v$  as:
 \begin{align}
    \Tilde{\Phi}_v(t) = \left( 1 + \alpha (t) \right) \Phi_v(t) + \beta(t) \, , \quad  \quad \forall t \in [t_1, t_2].
\end{align}
The FDI attacks described above can be represented by two variables: a scaling factor, $\alpha (t)$\,, and an additive factor, $\beta (t)$\,, which vary with different attack types, as present in Table\,\ref{tab1}.

\begin{table}[htbp]
\caption{Attack variables during different types of attacks, where $c$ denotes a real-valued constant, and $\Delta t=t-t_1$\,.} 
\begin{center}
\begin{tabular}{|c|c|c|}
\hline
\multirow{2}{*}{\textbf{Attack Types}}&\multicolumn{2}{|c|}{\textbf{Attack variables}} \\
\cline{2-3} 
& scaling factor, $\alpha (t)$ & additive factor, $\beta (t)$ \\
\hline\hline
Poisoning & 0 &  $ \sim \mathcal{N}(\mu_C, \sigma_C^2)$ \\
\hline
DoS & 0 &  $\Phi_{v}(t_1) - \Phi_v(t)$ \\
\hline
Step & $c$ &  0 \\
\hline
Ramp &  \,\,\,\,$ c \, \Delta  t$ \,\,\,\,&  0 \\
\hline
RTW & $ c \, \Delta  t$ &  $ c \, \Phi_{v}(t_1)\, \Delta  t $ \\
\hline
\end{tabular}
\label{tab1}
\end{center}
\end{table}

\subsection{Quantitative Comparison and Interpretation} 
\begin{figure}[h]
    \centering
    \includegraphics[width=\linewidth]{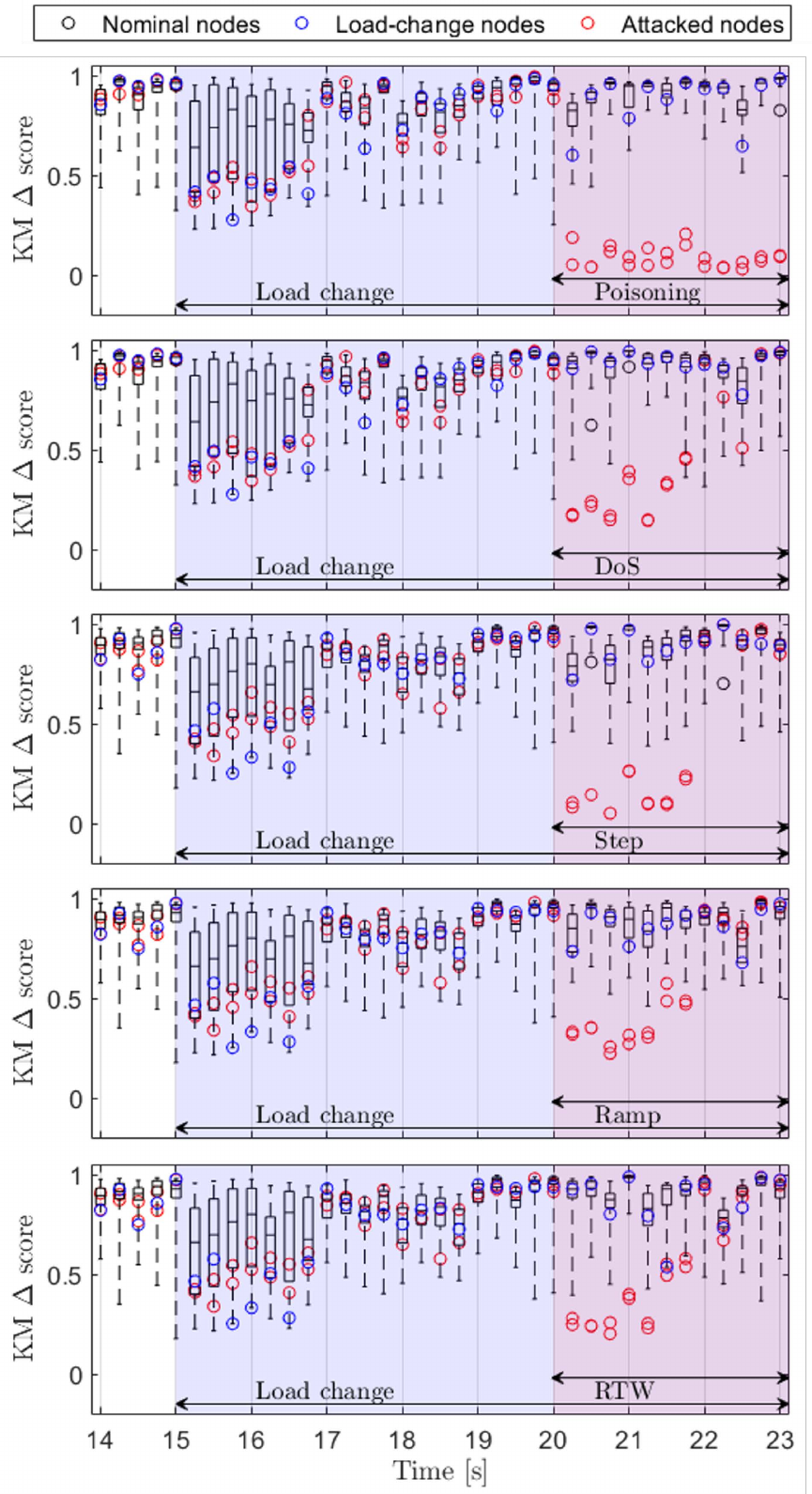}
    \caption{Quantitative comparison among different attacks}
    \label{fig:attack}
    \vspace{-0.25in}
\end{figure}
In this section, we present our findings for the 
attack scenarios described in the previous sections. For all the scenarios, we choose the learning window of 240 time steps ($12$s) and the prediction window of 40 time steps ($2$s) with a moving window of duration $0.25$s.
% Next, we obtain the error sequence corresponding to the prediction window and obtain the Koopman Modes of the error sequence. Finally, we calculate the desired distance measure from the approximate probability mass distribution of the Koopman modes. We repeat the analysis on  moving window of measurements. 
% Thus, we first utilize the first 12 second data to to learn the system and predict measurements for $13^{th}$ to $14^{th}$ second and repeat the steps with a moving window of 0.25 second. In other other words, next we consider the learning window of $0.25^{th}$ to $12.25^{th}$ second and the prediction window of $13.24^{th}$ to $14.25^{th}$ second and so on. 
Our objective is to visualize the embedding of the attack signature on the KMs and to quantitatively compare the results among different attack scenarios. Thus, we utilize the attack knowledge to visually capture the attack signature embedded in the KMs. The plots corresponding to each attack scenario in Fig. \ref{fig:attack} show our analysis from $14-23$s. Here, a large load change initiates at $15$s and the attack starts from $20$s. Now, let us compare our findings for the attack scenarios to better comprehend the influence of the attack and natural events on the Koopman modes. % and only the nominal scenario exhibits our analysis from $14^{th}$ second to $23^{rd}$ second.

\textbf{Grid-event vs. Attack:} \textit{Key observation: The attacked nodes distinctively appear as outliers after an attack.} 
In all the scenarios studied in this work, during and after the load change event, the origin bus nodes of the event do not appear as outliers. Rather, the whole network responds to such a natural grid-event, and thus a change in overall KM $\Delta$-score distribution with a larger spread is noticeable after $15$s in all scenarios.  On the other hand, after the attack occurrence, only the KMs corresponding to the attacked nodes capture the attack signature as a result the attacked nodes appear as outliers with a lower KM $\Delta$-score.  Such emergence of outliers with a lower KM $\Delta$-score indicates the atypical larger distance between attacked nodes and nominal nodes. However, the load-change nodes exhibit a closer connection with all other nodes throughout the event duration. Most importantly, regardless of the presence of both the grid event and the attack, only the attacked nodes emerge as outliers. Thus, the proposed algorithm can capture the attack signature even in the presence of the transient system dynamics arising due to grid events. The output of this algorithm, along with the findings, will be utilized to generate an automated detection algorithm in our future work. 

\textbf{Specific attacks:} \textit{Key observation: The range of the KM $\Delta$-score of the outliers varies with  different attack.} 
It is evident from Fig. \ref{fig:attack} that the KM $\Delta$-score of the outliers varies with the different attack scenarios. For example, comparatively lower KM $\Delta$-scores of the outlier, i.\,e., attacked nodes in poisoning and step attack point out the larger distance between the attacked nodes and the nominal nodes. At the same time, during DoS, ramp, and RTW attacks, the outlier exhibits a higher KM $\Delta$-score and thus, a lesser distance between the attacked nodes and the nominal nodes. Such a lesser distance between the outliers and the nominal nodes may increase the difficulty in automatic detection without any prior attack knowledge. In such scenarios, the KM $\Delta$-score may not provide sufficient information to ensure reliable detection. Thus, the ease and reliability of the attack detection varies with the attack nature and this knowledge needs to be considered while generating the detection algorithm. Furthermore, such difference in the range of  KM $\Delta$-score with different attacks may capture latent information that can help in identifying the specific nature of the attack.

\textbf{Detectability window:} \textit{Key observation: The outliers emerge temporarily and rapidly dissipate into the nominal nodes with time.}
Fig. \ref{fig:attack} shows that the outliers are absorbed within the nominal nodes with time in all the scenarios, except the poisoning attack. Considering this fact,  we infer that it is essential to detect the presence of the attack as early as possible. Otherwise, as the attack propagates throughout the network due to the closed-loop control structure of the system, as well as the attack signature spatially disperses through the overall KMs distribution. Hence, it may become exceedingly difficult to detect the presence of an attack with time. Such knowledge regarding time constraints on attack detectability is imperative to detection algorithm generation.

\subsection{Robustness}
To ensure the robustness of the algorithm we evaluated the performance of the algorithm with varying parameters. For instance, we compare the results from using different data-driven approaches to implement the Koopman operator such as Dynamic Mode Decomposition (DMD), Arnoldi method, and Hankel DMD. We also monitored the results with varying learning, prediction, and moving windows. Moreover, in each of these cases, we employed both KL and JS divergence to compare the results. The proposed algorithm provides consistent results with all these varying parameters. % and thus, exhibits robustness 

\section{Conclusion and Future Work} 
This work focused on quantitative comparison and visualization of the event signature embedding on the KMs. Our results show that the signature of any physical disruptions or grid-event is observed across the network nodes. However, signatures of the cyberattacks are distinctly observed at the compromised nodes. Moreover, the proposed algorithm can capture this distinctive attack signature even during grid events with transient system dynamics. The work provides a quantitative comparison of the performance under different attacks as well and exhibits potential usage towards isolating the nature of attacks. Our results visually show the time constraints on detection owing to the attack propagation throughout the closed-loop network system with time. In this work, we also conducted an extensive study to ensure the robustness of the proposed algorithm regardless of the implementation methods, time windows, and choice of probabilistic distance measures. This paper provides the baseline for our future work which will focus on automated cyberattack detection and the development of an identification algorithm.

\bibliographystyle{IEEEtran}
\bibliography{ref}

\end{document}